# Particle-in-cell Modeling of Electron Beam Generated Plasma

Shahid Rauf,[1,4] D. Sydorenko,[2] S. Jubin,[3] W. Villafana,[3] S. Ethier,[3] A. Khrabrov,[3] and I. Kaganovich[3]

[1]Applied Materials, Inc., 3333 Scott Blvd., Santa Clara, California 95054, USA

[2]University of Alberta, Edmonton, Alberta T6G 2E9, Canada

[3]Princeton Plasma Physics Laboratory, 100 Stellarator Rd., Princeton, New Jersey 08543, USA

[4]shahid_rauf@amat.com

## Abstract

Plasmas generated using energetic electron beams are well known for their low electron temperature ($T_e$) and plasma potential, which makes them attractive for atomic-precision plasma processing applications such as atomic layer etch and deposition. A 2d3v particle-in-cell (PIC) model for an electron beam-generated plasma in Argon confined by a constant applied magnetic field is described in this article. Plasma production primarily occurs in the path of the beam electrons in the center of the chamber. The resulting plasma spreads out in the chamber through non-ambipolar diffusion with a short-circuit effect allowing unequal electron and ion fluxes to different regions of the bounding conductive chamber walls. The cross-field transport of the electrons (and thus the steady-state characteristics of the plasma) are strongly impacted by the magnetic field. $T_e$ is anisotropic in the electron beam region, but low and isotropic away from the plasma production zone. The plasma density increases, and the plasma becomes more confined near the region of production when the magnetic field strengthens. The magnetic field reduces both electron physical and energy transport perpendicular to the magnetic field. $T_e$ is uniform along the magnetic field lines and slowly decreases perpendicular to it. Electrons are less energetic in the sheath regions where the sheath electric field repels and confines the low-energy electrons from the bulk plasma. Even though electron and ion densities are similar in the bulk plasma due to quasi-neutrality, electron and ion fluxes on the grounded chamber walls are unequal at most locations. Electron confinement by the magnetic field weakens with increasing pressure, and the plasma spread out farther from the electron beam region.

## I. INTRODUCTION

Sub-nm scale precision is increasingly required in many critical plasma processing applications used for fabricating leading-edge microelectronics devices [1, 2]. Precise control of ion energy and ion / radical composition are necessary during critical plasma processes to achieve the requisite atomic-scale fidelity. Two applications that will benefit from such precise control are atomic layer etch [3, 4] and atomic layer deposition [5], which attempt to remove or add the material atomic layer by atomic layer while leaving the underlying material undamaged. Previous studies have shown that ions with energy higher than a few eV can alter or damage the sub-surface material [2, 6, 7]. True atomic precision plasma processing is therefore difficult to achieve using radio frequency (RF) plasmas (e.g., inductively and capacitively coupled plasmas) where ion energies are generally greater than 10 eV. Electron beam-generated plasmas (hereafter referred to as ebeam plasmas) have been shown to have characteristics suitable for atomic precision plasma processing [8]. With high ion density, low electron temperature ($T_e$) and ion energies less than 5.0 eV in plasmas of molecular gases [9, 10], ebeam plasmas can deliver a large flux of low energy ions to adjacent substrates. Significant amount of research has been reported on the characterization of ebeam plasmas [9 - 12] as well as materials processing using these plasmas [2, 13 - 16]. Work has also been done on developing industrial scale ebeam plasma processing systems [17, 18].

Several authors have previously described models for magnetized ebeam plasmas in the literature. Fernsler *et al.* [19] developed model for a large area ebeam plasma and used this model to identify the plasma properties needed for the resulting sheet plasma to efficiently reflect microwaves. They correctly predicted that the plasma is generated more efficiently by energetic electron beam compared to RF, leading to a lower $T_e$ in ebeam plasmas. Lock *et al.* [20] described a global model for an $N_2$ ebeam plasma, which highlighted that gas dissociation is weaker and ion to excited state density ratio higher in the ebeam plasma compared to RF plasmas. Petrov and colleagues [11, 21, 22] have developed a model which combines on-axis solution of the Boltzmann equation for the electron energy distribution (EED) with a fluid model for the bulk plasma. These authors have examined control of the EED, the influence of magnetic fields on plasma properties, and the effects of $N_2$ and $SF_6$ addition to Ar on charged species concentrations and $T_e$. Petrov *et al.* also compared modeling results to experiments with generally good agreement observed. Rauf *et al.* [18] described a hybrid 3-dimensional (3D) model for ebeam plasmas where a Monte Carlo model was used for plasma generation by the electron beam. Levko and Raja examined the influence of $SF_6$ addition to Ar on the stability of electron beam driven discharge using a 1D PIC model [23]. Huang *et al.* described a 1D PIC model for an electron beam plasma [24]. They found that the electron energy distribution is Druyvesteyn-like with a high energy tail.

Most of the multi-dimensional models for electron beam plasmas are fluid-based and use fluid assumptions about charged species transport in the magnetized plasma. Classical transport equations describing the effect of magnetic field on electron transport coefficients [25] are often used in these models. The work in Ref. [18] however illustrates that such models need significant adjustments to be able to match experiments. Plasma physical and energy transport in this low-pressure magnetized plasma appears to be non-classical. To self-consistently understand the characteristics of ebeam plasma, we simulated them using a particle-in-cell (PIC) based kinetic model. The PIC modeling results are discussed in this article, which are currently being theoretically analyzed to understand electron transport and develop models for electron transport coefficients. This detailed physics analysis will be discussed in a subsequent publication.

This article is organized in the following manner. The computational model is described in Sec. 2. Modeling results are discussed in Sec. 3 and Sec. 4 includes a summary.

## II. COMPUTATIONAL MODEL

The simulations in this article have been done using EDIPIC, a 2-dimensional (2d3v) electrostatic particle-in-cell (PIC) modeling code. EDIPIC is open source and available on GitHub with the necessary documentation [26]. It uses a standard explicit leap-frog algorithm in Cartesian geometry, with the Boris scheme for particle advance [27]. The electrostatic field is obtained from the Poisson's equation solved using the PETSc library [28]. The code includes a Monte-Carlo model of elastic, inelastic, and ionization electron-neutral collisions. Cross-sections for collisions with Argon neutrals used in the simulations described below are from Ref. [29]. EDIPIC can simulate crucial atomistic and plasma-surface interaction processes needed for simulations of partially ionized plasmas, including, but not limited to, the secondary electron emission induced by electrons and ions. The Langevin model of Coulomb collisions for electrons is also implemented. EDIPIC has been verified in several international benchmarks [30, 31].

EDIPIC is written in Fortran 90 and parallelized using Message Passing Interface (MPI). Good scalability of up to 400 CPU cores has been demonstrated. The code is equipped with numerous, diverse diagnostic capabilities, including but not limited to, the phase-space data and ion and electron velocity distribution functions, as well as spectral analysis procedures required to study wave propagation in plasmas.

Electron temperature can be anisotropic in the plasma, in particular along the path of the electron beam. We, therefore, compute the components of the temperature along each coordinate direction using [32]:

$$\boldsymbol{T}_{s:x,y,z} = \boldsymbol{m}_s\langle \boldsymbol{c}^2_{s:x,y,z}\rangle = \boldsymbol{m}_s\left(\langle \boldsymbol{v}^2_{s:x,y,z}\rangle - \langle \boldsymbol{v}_{s:x,y,z}\rangle^2\right),$$

where the subscript $s$ denotes species, $m_s$ is the mass of the particle, $v_{s;x,y,z}$ is the velocity in the $x$, $y$, and $z$ directions, $\vec{\boldsymbol{c}}_s = \vec{\boldsymbol{v}}_s - \langle\vec{\boldsymbol{v}}_s\rangle$ is the particle random velocity, and angular brackets $\langle\cdot\rangle$ denote averaging over particles. We also calculate components of the heat flow vector defined as [33]:

$$\boldsymbol{Q}_{s:x,y,z} = \boldsymbol{n}_s\boldsymbol{m}_s\langle \boldsymbol{c}^2_s c_{s:x,y,z}\rangle,$$

where $\boldsymbol{n}_s$ is the density of species *s*. These quantities are used in the discussion below.

## III. COMPUTATIONAL RESULTS

2-dimensional simulations of the electron beam generated plasma (ebeam plasma) have been done in Cartesian geometry for the plasma reactor configuration shown in Fig. 1. As the plasma system is symmetric around *x* = 0, only ½ of the plasma is simulated with appropriate symmetry boundary conditions imposed at *x* = 0. 2 keV electrons are launched in the *z*-direction from a 2.8 mm wide perforated metal window at the bottom. The electron beam width and current quoted in this paper are for ½ of the window from 0 – 2.8 mm. All surfaces of the plasma chamber are assumed grounded. A *z*-directed uniform magnetic field is applied across the plasma. All simulations have been done for 0.8 ms, and the plasma properties reach steady state in this time for the conditions considered.

Before we examine the effect of magnetic field and gas pressure on the characteristics of the ebeam plasma, we first describe plasma properties for the following operating conditions: 20 mT gas pressure in Ar, 100 G magnetic field, 12.5 mA/m electron beam current (1/2 width), and 2 keV beam electron energy. The steady-state plasma potential, source of electrons and $Ar^+$ ions, electron density ($n_e$), and electron temperature ($T_{ex}$) are shown in Fig. 2. The beam electrons are launched in the $z$-direction and the applied $z$-directed magnetic field keeps them well confined near the center of the chamber. Consequently, plasma production primarily occurs in a narrow region near the chamber center. The beam electrons lose some energy along their trajectory as they collide with the background gas. The slower electrons near the top wall ionize the background gas at a slightly higher rate due to the higher ionization cross-section at lower energy. The electrons and ions generated during the ionization collisions spread out fairly symmetrically around $z = 4.5$ cm, despite the slightly skewed source. The steady-state electron density $n_e$ (and consequently the ion density since quasi-neutrality is maintained) has the typical diffusion-like profile with a peak at the chamber center and lower density near the walls. However, as discussed later in the article, the magnetic field reduces electron transport perpendicular to the magnetic field, which significantly impacts the density profile. The electron velocity distribution function (EVDF) is anisotropic in the beam region, and we have computed the components of $T_e$ in the different directions separately. Outside of the beam region (x > 2.8 mm), $T_{ex} \sim T_{ey} \sim T_{ez}$. However, in the path of the energetic electron beam, the electron temperature is anisotropic with $T_{ez} > T_{ex}$. We have plotted the x-component of electron temperature $T_{ex}$ in Fig. 2(c). The electron temperature is more uniform both along and across the magnetic field lines compared to $n_e$, so electron energy spreads efficiently at these low-pressure conditions. $T_{ex}$ is uniform in the bulk plasma along the magnetic field axis but decreases slowly in the $x$-direction. The electrons are less energetic near all surfaces where the sheath electric field repels the low-energy electrons from the bulk plasma and keeps them confined. The plasma potential is 4-5 times $T_{ex}$, and is not as uniform in the bulk plasma as $T_{ex}$.

We next consider the effect of applied magnetic field strength on plasma properties at 20 mTorr. For all these simulations, beam electrons are launched with 2 keV energy and the beam current launched from the beam inlet is 12.5 mA/m (1/2 width). The magnetic field is varied between 50 – 200 G. The effect of magnetic field strength on steady state $n_e$ and $T_{ex}$ are shown in Figs. 3 and 4, respectively. While not shown in this article, the electron source does not change significantly with increasing magnetic field. This is expected as the plasma is primarily produced by the beam electrons, which remain confined for the range of magnetic field strength considered. As the magnetic field is increased, the gradient of electron density perpendicular to the magnetic field increases. This is due to reduced cross-field transport of electrons compared to electron transport along magnetic field lines. As a result of better electron confinement, the maximum value of $n_e$ is found to increase with magnetic field strength. We will look closely at the profile of $n_e$ perpendicular to the electron beam in Fig. 5. As discussed earlier, $T_{ex}$ is mostly uniform in the $z$-direction except near the boundaries where the kinetic energy of the trapped electrons is converted to potential energy as they are repelled by the electric field in the sheath region. Peak $T_{ex}$ increases slightly with increasing magnetic field and the gradient of $T_{ex}$ across the magnetic field increases slightly as well, which is explained later using the results in Fig. 6.

To quantitatively understand how $n_e$ and $T_{ex}$ are impacted by the magnetic field, we have plotted $n_e$ and $T_{ex}$ at mid-height (averaged between $z$ = 3.5 – 5.5 cm) as a function of $x$ in Fig. 5. Both the absolute and normalized $n_e$ are shown. Peak $n_e$ clearly increases and $n_e$ decays faster in the $x$-direction with increasing magnetic field. Also, as mentioned earlier, peak $T_{ex}$ increases slightly with magnetic field. Although difficult to discern in Fig. 4, $T_{ex}$ decays faster in the $x$-direction when the magnetic field is stronger. It is clear from the results in Figs. 3 – 5 that electron physical transport is more strongly impacted by the magnetic field than electron energy transport.

Although the electrons collide with the background gas as they spread in the chamber, they lose relatively little energy in these collisions due to the mass difference between electrons and neutral Argon atoms. Consequently, the electron temperature drops very gradually across the magnetic field.

To help further explain the trends of $n_e$ and $T_{xe}$ with the magnetic field, we have plotted the $x$-directed mean electron velocity $v_{ex}$ and heat flux $Q_{xe}$ in Figs. 6(a) and 6(b) as a function of the magnetic field at 20 mTorr. $v_{xe}$ and $Q_{xe}$ have been averaged over $z = 3.5 – 5.5$ cm. The $z$-directed mean electron velocity $v_{ze}$ has been plotted in Fig. 6(c) as a function of the magnetic field. $v_{ze}$ is averaged from $x = 2.5 – 3.5$ cm and plotted as a function of $z$. There are no beam electrons in the region where $v_{ze}$ has been plotted in Fig. 6(c). As the magnetic field is increased, electrons are better confined, and their diffusion across the magnetic field lines is impeded. Therefore, $v_{xe}$ decreases with increasing magnetic field. Also due to better electron confinement, less electron energy leaves the beam region in the $x$-direction at the higher magnetic field. Heat flux also decreases more quickly in the $x$-direction at a stronger magnetic field as electron energy can escape more easily in the $z$-direction. Classical transport theory would indicate that electron transport should not be impacted by the magnetic field along the direction of the magnetic field. However, as shown in Fig. 6(c), $v_{ze}$ is a strong function of magnetic field and is lower in the bulk plasma at higher magnetic field. The $v_{ze}$ plot has been truncated at ±240 m/s to highlight the velocity in the plasma bulk. $v_{ze}$ is significantly larger in the sheath regions.

Even though equal number of electrons and ions leave the plasma region at the walls in steady state, their fluxes don't have to balance at each location on the wall as the walls are conductive. Thus, the ambipolar electric field which would result in ambipolar diffusion is effectively short-circuited [34]. To illustrate the effect of magnetic field on charged species transport out of the plasma, we have plotted the ion and electron fluxes at the top, bottom, and right walls in Fig. 7. These fluxes have been averaged over 5 cells adjacent to the walls (1.07 mm). At the top and bottom walls, the electron current is high in the beam regions due to the beam electron current, so we have not shown electron flux for $x < 2.8$ mm. Closer to the region of plasma production, more electrons are observed to leave at the walls compared to ions. As we move to larger $x$, the electron flux falls off more quickly due to electron confinement by the magnetic field. At a certain $x$-location, the ion flux surpasses the electron flux. This transition occurs at a smaller $x$ for larger magnetic field due to better electron confinement. More electrons are observed to leave at the top surface compared to the bottom, particularly at smaller $x$ near the region of plasma production. Although the ions are not magnetized for the magnetic field strengths considered, the ion density matches that of electrons throughout the bulk of the plasma to maintain quasineutrality. Only in the sheath regions near the walls do the ion and electron densities diverge. One can observe in Figs. 7(a) and 7(b) that the ion flux is higher near $x = 0$ for a larger magnetic field due to higher ion density. However, the ion flux decreases more quickly in the $x$-direction at the larger magnetic field. Virtually no electrons leave at the right wall for the conditions examined. Ions do exit there, but their flux is smaller on the sidewall compared to the top and bottom surfaces. Ion flux to the right wall is smaller at a larger magnetic field due to better plasma confinement.

Plasma steady-state characteristics are also sensitive to gas pressure due to its influence on collisional processes and charged species transport. We next examine the effect of varying the gas pressure from 10 – 40 mTorr with a fixed 200 G magnetic field. Other operating conditions are 6.25 mA/m beam electron current (1/2 width) and 2 keV beam electron energy. The steady-state electron density $n_e$ is shown in Fig. 8 for different pressures. As the neutral gas density increases proportional to the pressure, beam electrons collide more often with the background gas.

Consequently, more charged species are produced and $n_e$ increases with increasing pressure. Charged species transport (drift and diffusion) is also slower along the magnetic field lines at higher pressure, which further helps increase $n_e$. The electron density $n_e$ decreases faster in the $x$-direction at lower pressure due to better confinement by the magnetic field, which can be seen in Fig. 10(b).

The effect of pressure on steady-state $T_{ex}$ is shown in Fig. 9. We comment on the effect of pressure on the anisotropy of electron temperature when the results in Figs. 10 and 11 have been presented. However, peak $T_{ex}$ increases considerably with pressure. For the range of gas pressures modeled, $T_{ex}$ is reasonably uniform in the $z$-direction except for a sharp drop near the boundaries where the kinetic energy of the trapped electrons is converted to potential energy as they are repelled by the sheath electric field.

To quantitatively examine the effect of gas pressure on $n_e$ and $T_{ex}$, we have plotted $n_e$ and $T_{ex}$ at mid-height (averaged between $z$ = 3.5 – 5.5 cm) as a function of $x$ in Fig. 10. Both the absolute and normalized $n_e$ are shown. The electron density increases with pressure, but the gradient of the electron density is shallower at higher gas pressure. At the lowest pressure considered, 10mTorr, the density is peaked sharply within 1 cm of the beam, plateauing at a lower value across the remaining distance. The trends vs. pressure are related to more intense plasma production and enhanced electron diffusion across the magnetic field lines at higher pressures. The peak $T_{ex}$ increases considerably with gas pressure, but gas pressure only has a minor impact on the gradient of $T_{ex}$.

To help explain the trends of $n_e$ and $T_{xe}$ with gas pressure, we have plotted the electron velocity and $z$-component of the electron temperature $T_{ze}$ in Fig. 11 as a function of gas pressure at 200 G magnetic field. In Figs. 11(a) – (c), the velocity and temperature have been averaged over $z$ = 3.5 – 5.5 cm and plotted as a function of $x$. In Fig. 11(d), $v_{ze}$ has been averaged from $x$ = 2.5 – 3.5 cm and plotted as a function of $z$. This $v_{ze}$ vs. $z$ plot is outside the beam region where electrons are not being produced. Because of their low density, contribution of high energy beam electrons to the reported velocity and temperature is negligible. As the beam electrons have high velocity in the $z$-direction, the velocity of newly created electrons and scattered beam electrons is initially biased in the $z$-direction. Subsequent collisions generally increase the $x$-component of the velocity. The electrons collide more frequently with the background gas at higher gas pressure, so $v_{xe}$ increases with increasing gas pressure. This increase in $v_{xe}$ with pressure comes at the expense of $v_{ze}$, and as shown in Fig. 11(c), $v_{ze}$ is significantly lower in the beam region at higher pressure. Based on Figs. 10(c) and 11(b), even though $T_{ex} \sim T_{ez}$ in the region outside the beam, $T_e$ is anisotropic in the beam region. Electrons get heated in the beam region when the beam electrons collide with the background gas. Heating in the initial collisions is biased in the $z$-direction due to the $z$-directed beam electrons. As there are more collisions at 40 mTorr, more energy gets transferred in the $x$-direction and anisotropy in $T_e$ decreases. $v_{ze}$ in the bulk plasma is observed to increase with higher pressure (Fig. 11(d)) indicating that electrons transport is also altered along the magnetic field direction. The $v_{ze}$ plot has been truncated at ±240 m/s to highlight the velocity in the plasma bulk and $v_{ze}$ is significantly larger in the sheath regions.

To examine charged species' transport out of the plasma chamber, we have plotted the electron and ion fluxes to the walls in Fig. 12 as a function of pressure. Electron flux has not been shown in the beam region (x < 2.8 mm) as it is dominated by the beam electrons. We find that even though an equal number of electrons and ions leave the plasma in steady state, electron and ion fluxes are not equal at different locations on the wall. Near the beam region at both the top and bottom surfaces, the electron flux is higher than the ion flux as electrons are confined by the magnetic field and can leave more easily in the $z$-direction. Crossing a certain $x$ location, which increases with pressure, the ion flux becomes larger than the electron flux at both the top and bottom surfaces. On the sidewall, only ions exit, and the electron flux is negligible for the range of pressures examined.

## IV. CONCLUSIONS

Plasmas generated using energetic electron beams (ebeam plasmas) are well known in the literature for their low $T_e$ and plasma potential. The resulting low ion energy makes them attractive for plasma processing applications requiring atomic-scale precision, such as atomic layer etch and deposition. A 2-dimensional particle-in-cell (PIC) model for ebeam plasmas confined using a constant magnetic field is described in this article. It is found that plasma production primarily occurs in the path of the magnetically confined beam electrons. The resulting plasma spreads out in the chamber through separate mechanisms for electrons and ions, as the ambipolar electric field is short-circuited by the conducting chamber walls. Plasma transport is strongly impacted by the magnetic field. $T_e$ is observed to be anisotropic ($T_{ez} > T_{ex}$) in the beam region, but it is low and isotropic away from the region of plasma production. Charged species densities increase and become better confined near the plasma production region as the magnetic field becomes stronger. The magnetic field reduces both electron physical and energy transport perpendicular to the magnetic field. $T_e$ is uniform along the magnetic field lines and slowly decreases perpendicular to it. Electrons are less energetic in the sheath regions, where the sheath electric field pushes back the low energy electrons from the plasma. Even though electron and ion densities are similar in the bulk plasma due to quasi-neutrality, electron and ion fluxes on the grounded chamber walls are unequal at most locations. Electron flux exceeds the ion flux near the region of plasma production.

However, the electrons are better confined by the magnetic field and electron flux drops quickly away from the electron beam area. The ions are observed to leave on the sidewall too, where the electron flux is negligible. Electron confinement by the magnetic field weakens with increasing pressure, and the electrons spread out farther from the region of production. $T_e$ becomes less anisotropic in the region of production at higher gas pressure with more collisions. $T_e$ is observed to increase with increasing pressure, which is attributed to electron energy thermalizing better in the production region at higher pressure.

These PIC simulations have been done to self-consistently understand electron transport in the low-pressure magnetized ebeam plasmas. The modeling results are reported in this article. We are currently theoretically analyzing these results to understand charged species transport and how electron transport coefficients scale with magnetic field and gas pressure. Results from this theoretical analysis will be presented in an upcoming publication.

## V. AUTHOR DECLARATION

### Data Availability

The data that support the findings of this study are available from the corresponding author upon reasonable request.

### Conflict of interest

The authors have no conflicts to disclose.

## Figure Captions

1. Geometry used in the plasma simulations.
2. Steady-state (a) electron production rate $S_e$, (b) electron density $n_e$, (c) electron temperature $T_e$, and (d) potential. These results are at 20 mTorr gas pressure, 100 G magnetic field, 12.5 mA/m beam electron current, and 2 keV beam electron energy.
3. Steady-state electron density $n_e$ at 20 mTorr gas pressure for (a) 50, (b) 100, (c) 150 and (d) 200 G magnetic field. Other operating conditions are 12.5 mA/m beam electron current, and 2 keV beam electron energy.
4. Steady-state electron temperature $T_e$ at 20 mTorr gas pressure for (a) 50, (b) 100, (c) 150 and (d) 200 G magnetic field. Other operating conditions are 12.5 mA/m beam electron current, and 2 keV beam electron energy.
5. The effect of magnetic field on (a) electron density $n_e$ (b) normalized $n_e$, and (c) electron temperature $T_e$ at the mid-plane. Operating conditions in the Ar plasma are 20 mTorr gas pressure, 12.5 mA/m beam electron current, and 2 keV beam electron energy.
6. The effect of magnetic field on (a) $x$-directed velocity of electrons $v_{xe}$ (b) electron heat flux in the $x$-direction $Q_{xe}$, and (c) $z$-directed velocity of electrons $v_{ze}$. Operating conditions in the Ar plasma are 20 mTorr gas pressure, 12.5 mA/m beam electron current, and 2 keV beam electron energy.
7. The effect of magnetic field on the fluxes of electron and $Ar^+$ ions to the (a) top (b) bottom, and (c) right surfaces. Operating conditions in the Ar plasma are 20 mTorr gas pressure, 12.5 mA/m beam electron current, and 2 keV beam electron energy.

8. Steady-state electron density $n_e$ at 200 G magnetic field for (a) 10, (b) 20, (c) 30 and (d) 40 mTorr gas pressure. Other operating conditions are 6.25 mA/m beam electron current and 2 keV beam electron energy.
9. Steady-state electron temperature $T_e$ at 200 G magnetic field for (a) 10, (b) 20, (c) 30 and (d) 40 mTorr gas pressure. Other operating conditions are 6.25 mA/m beam electron current and 2 keV beam electron energy.
10. The effect of gas pressure on (a) electron density $n_e$ (b) normalized $n_e$, and (c) electron temperature $T_e$ at the mid-plane. Operating conditions in the Ar plasma are 200 G magnetic field, 6.25 mA/m beam electron current, and 2 keV beam electron energy.
11. The effect of gas pressure on (a) $x$-directed velocity of electrons $v_{xe}$, (b) $z$-directed temperature of electrons $T_{ze}$, (c) $z$-directed velocity of electrons $v_{ze}$ plotted as a function of $x$. (d) $z$-directed velocity of electrons $v_{ze}$ plotted as a function of $z$. Operating conditions in the Ar plasma are 200 G magnetic field, 6.25 mA/m beam electron current, and 2 keV beam electron energy.
12. The effect of gas pressure on fluxes of electron and $Ar^+$ ions to the (a) top (b) bottom, and (c) right surfaces. Operating conditions in the Ar plasma are 200 G magnetic field, 6.25 mA/m beam electron current, and 2 keV beam electron energy.

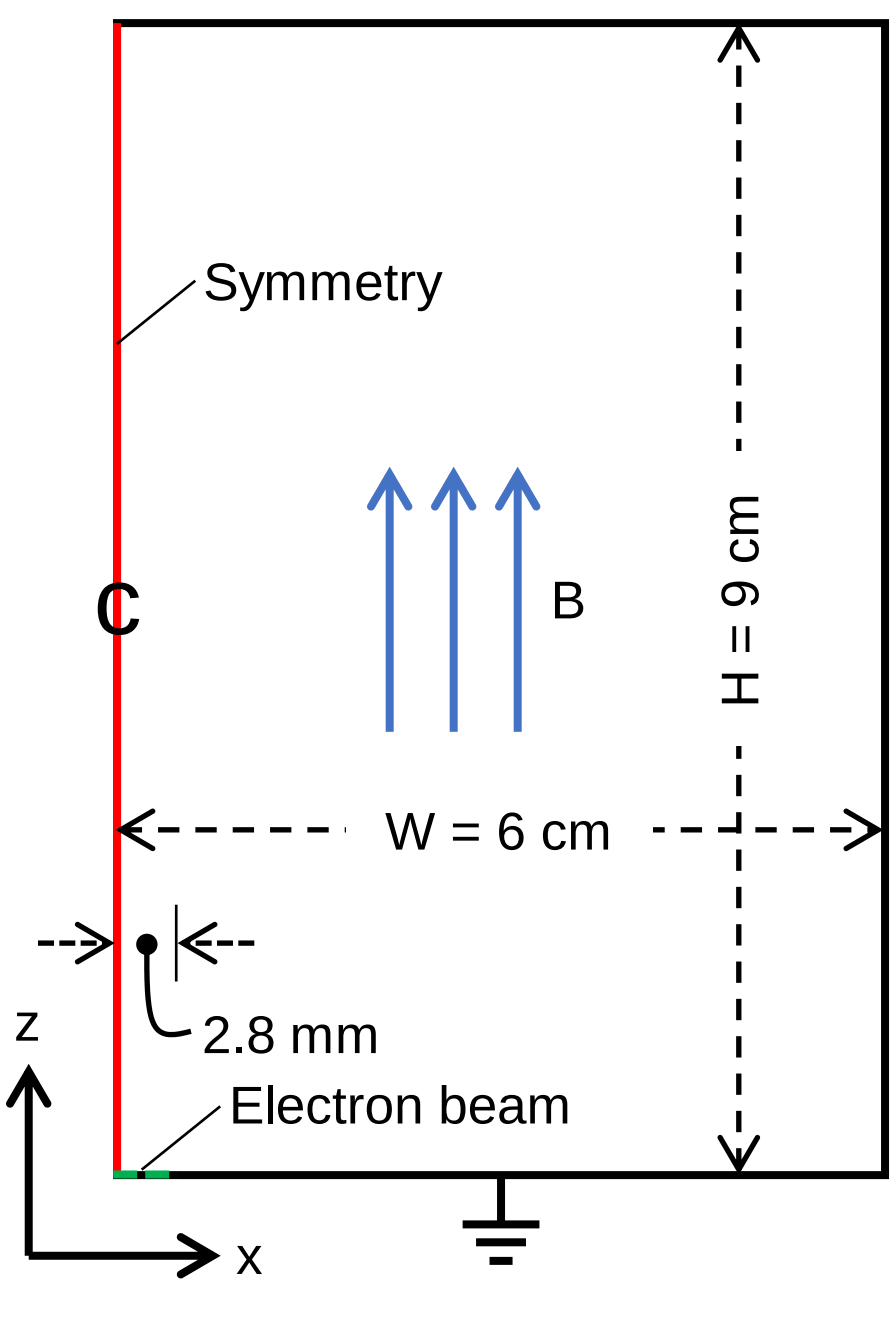

Figure 1
Rauf *et al.*

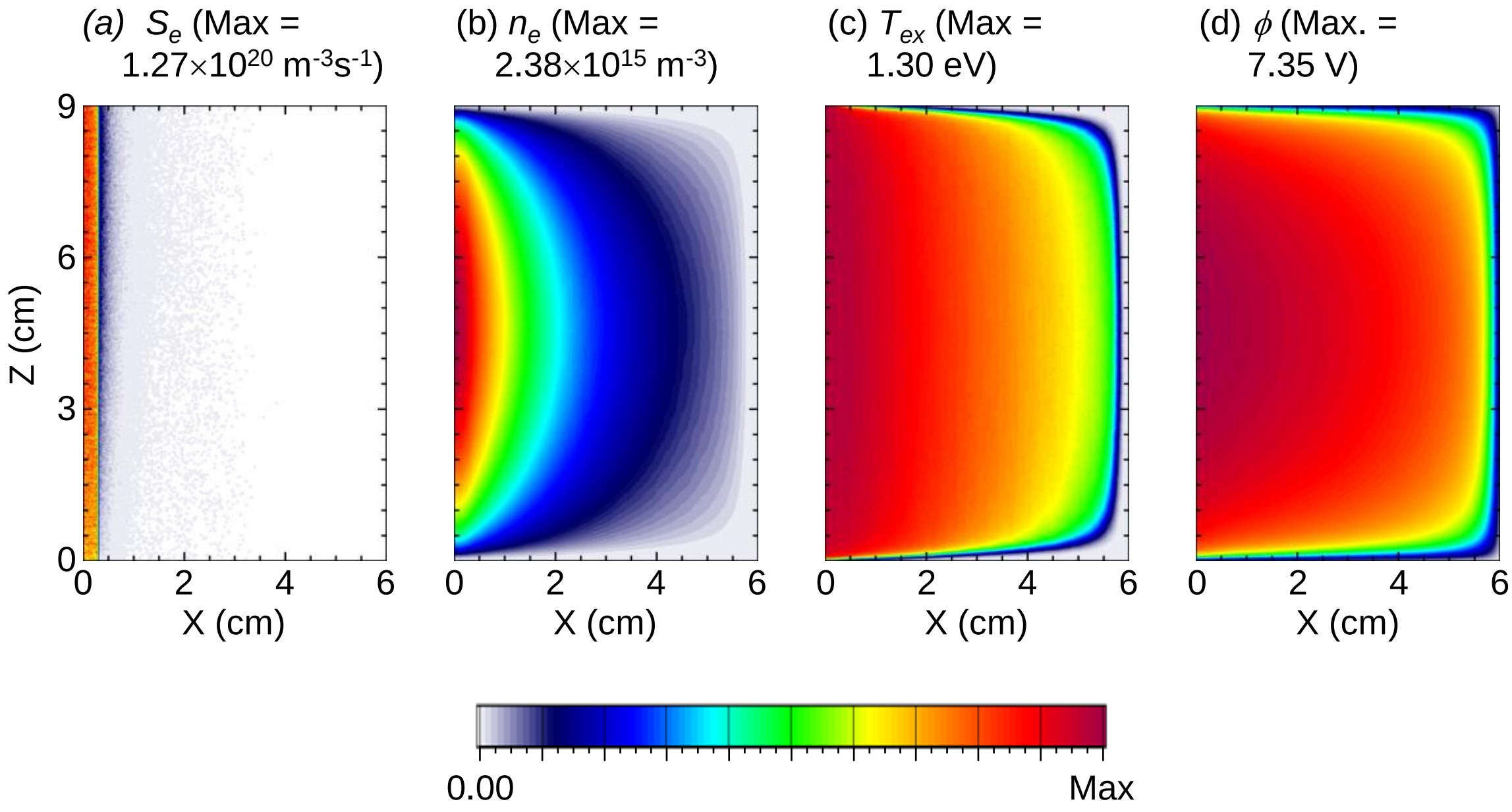

Figure 2
Rauf *et al.*

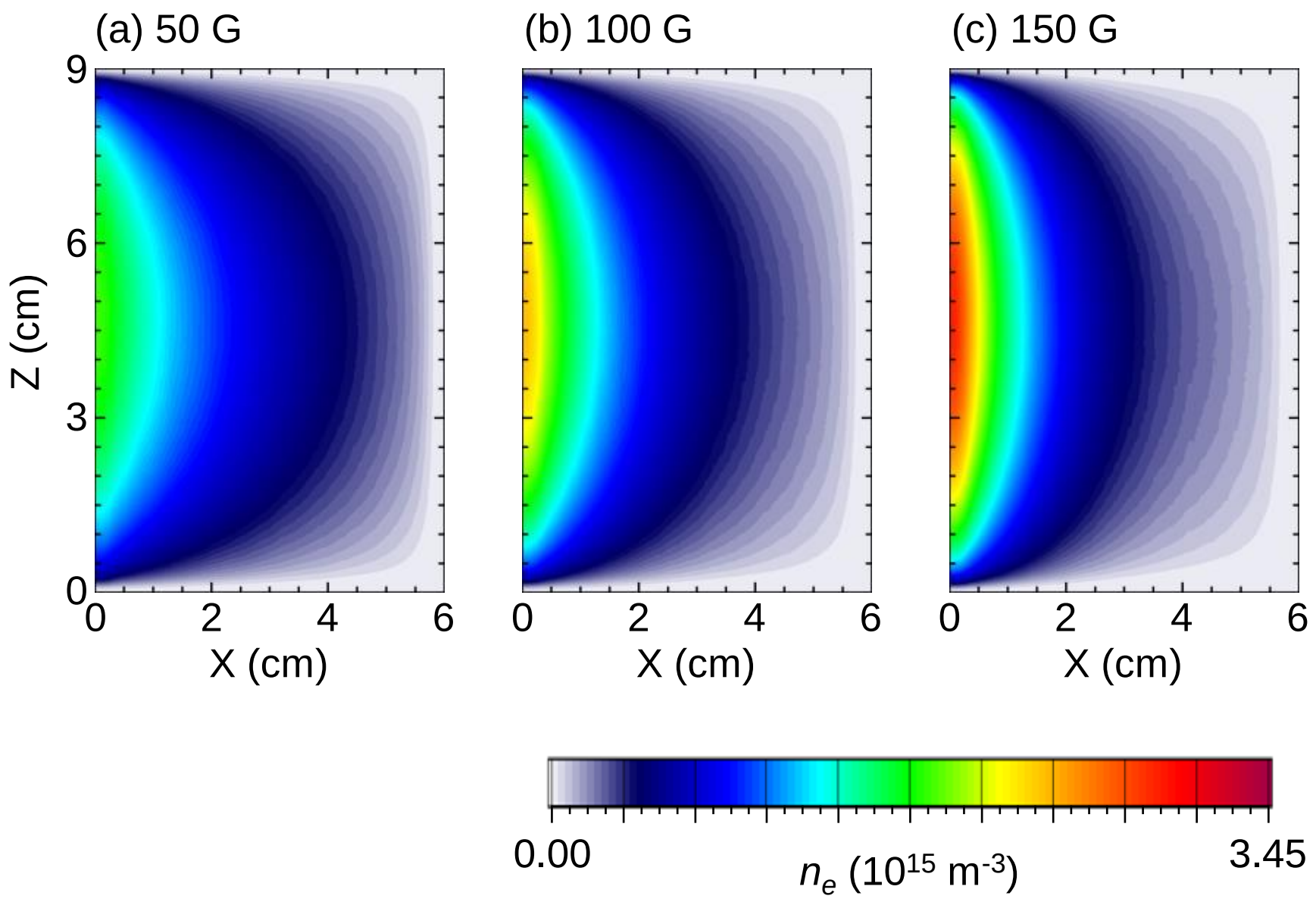

Figure 3
Rauf *et al.*

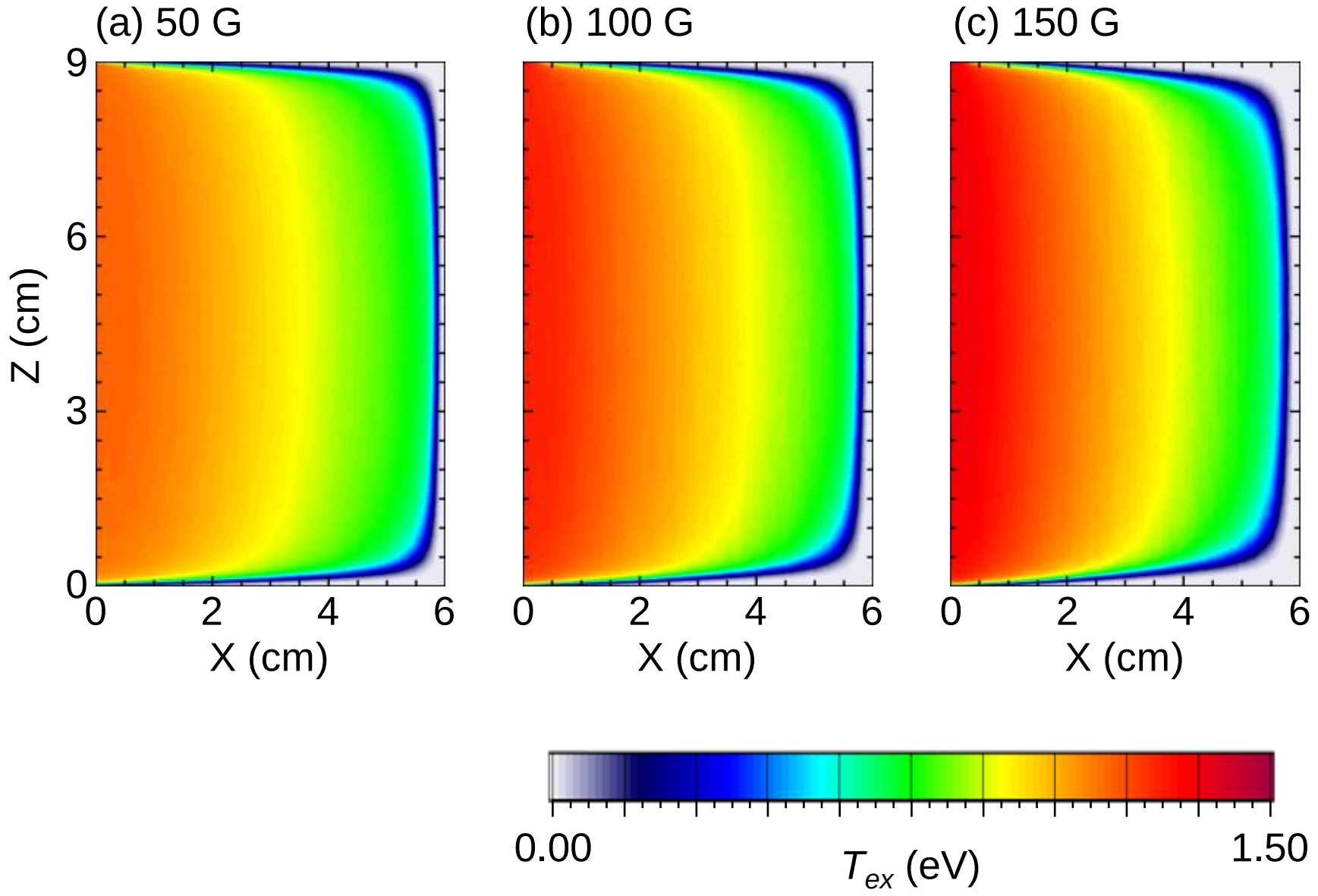

Figure 4
Rauf *et al.*

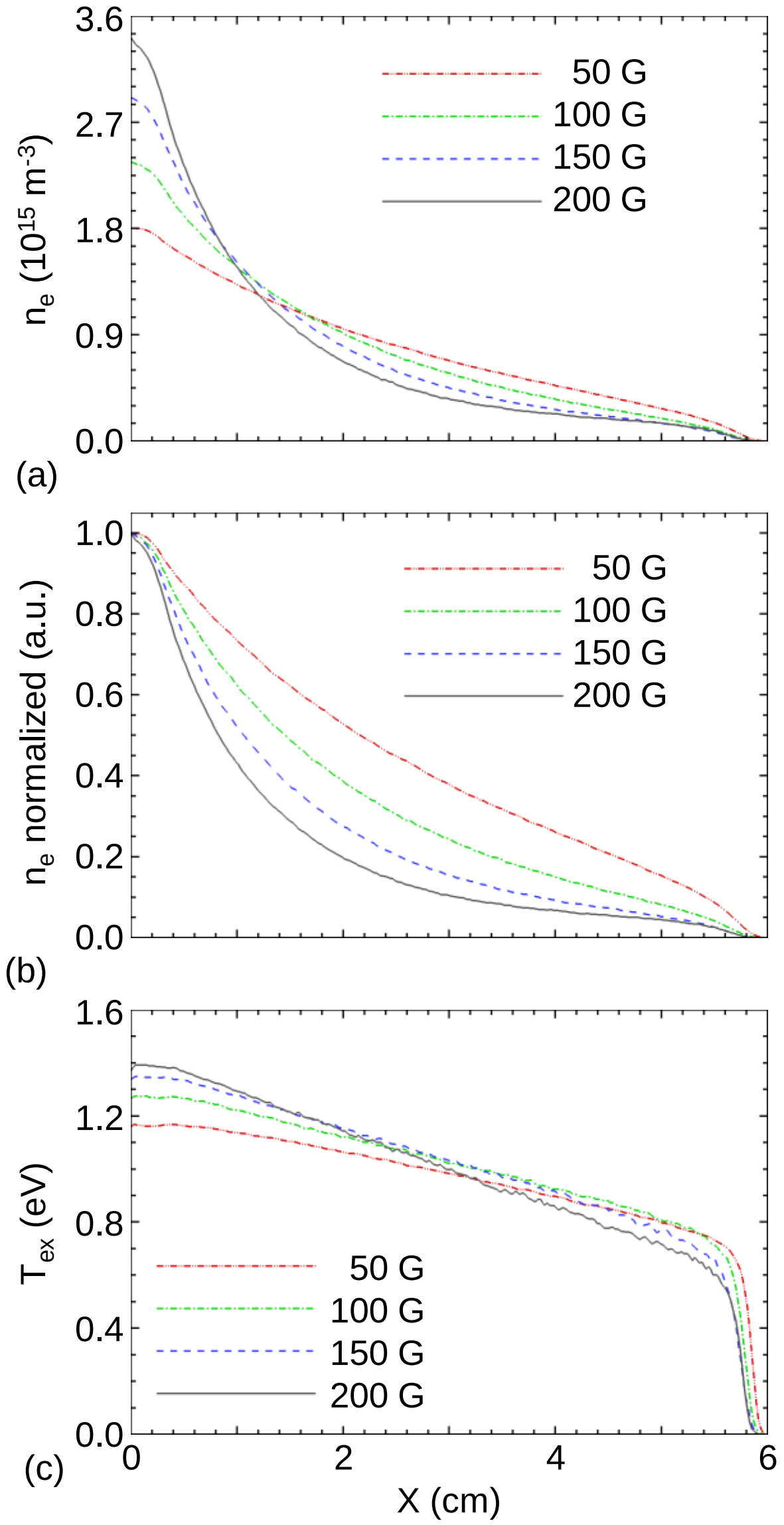

Figure 5
Rauf *et al.*

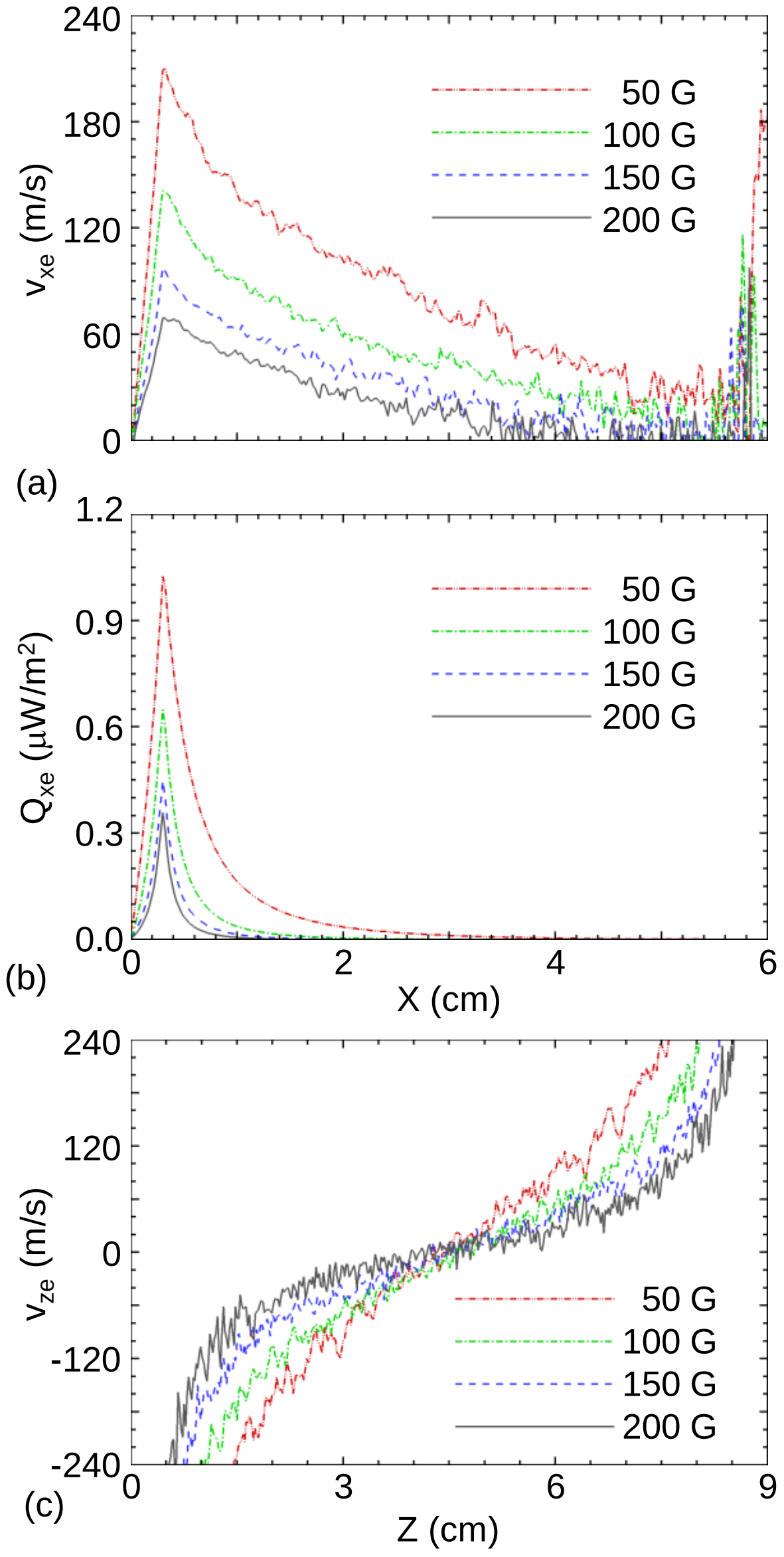

Figure 6
Rauf *et al.*

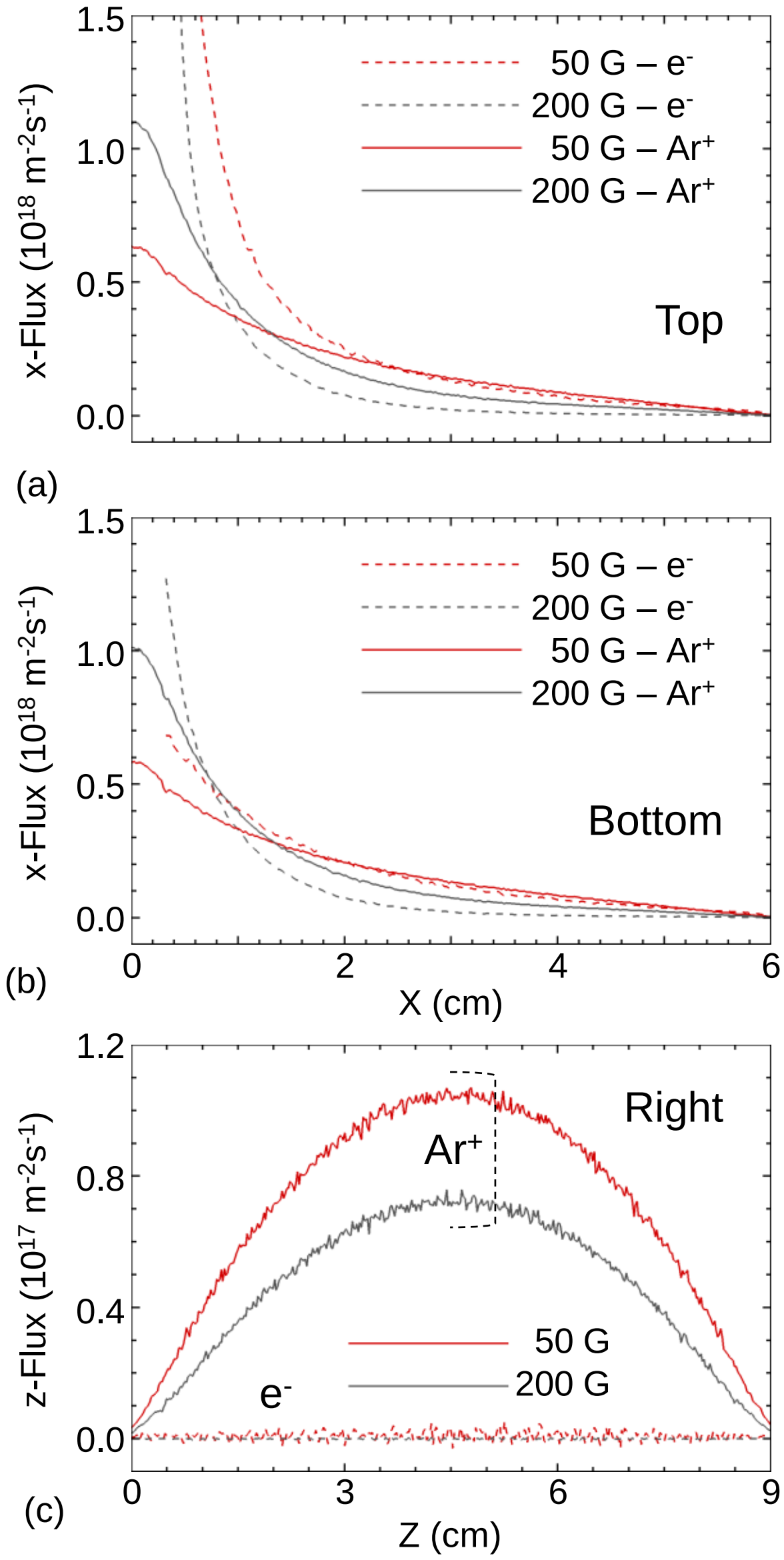

Figure 7
Rauf *et al.*

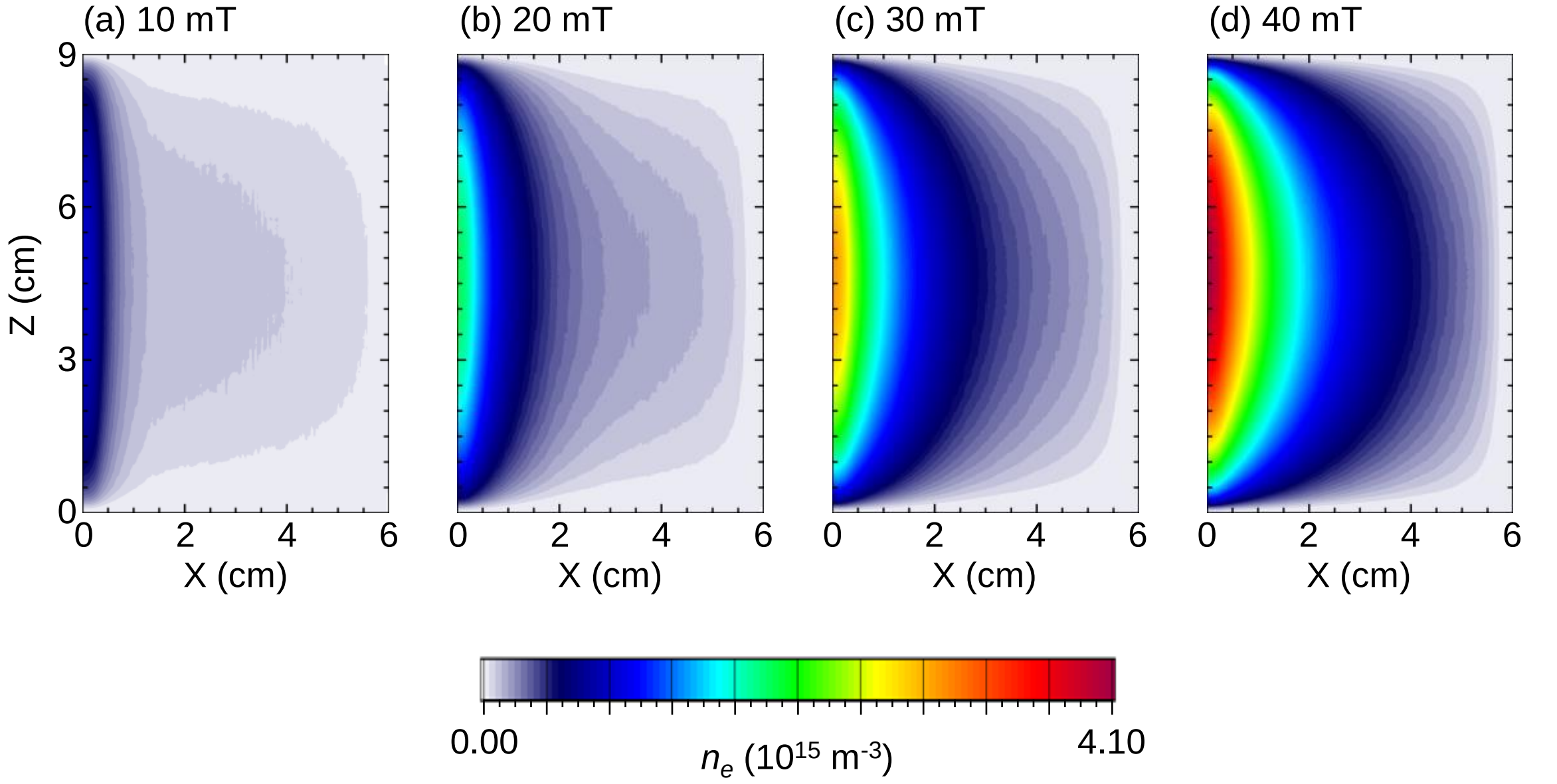

Figure 8
Rauf *et al.*

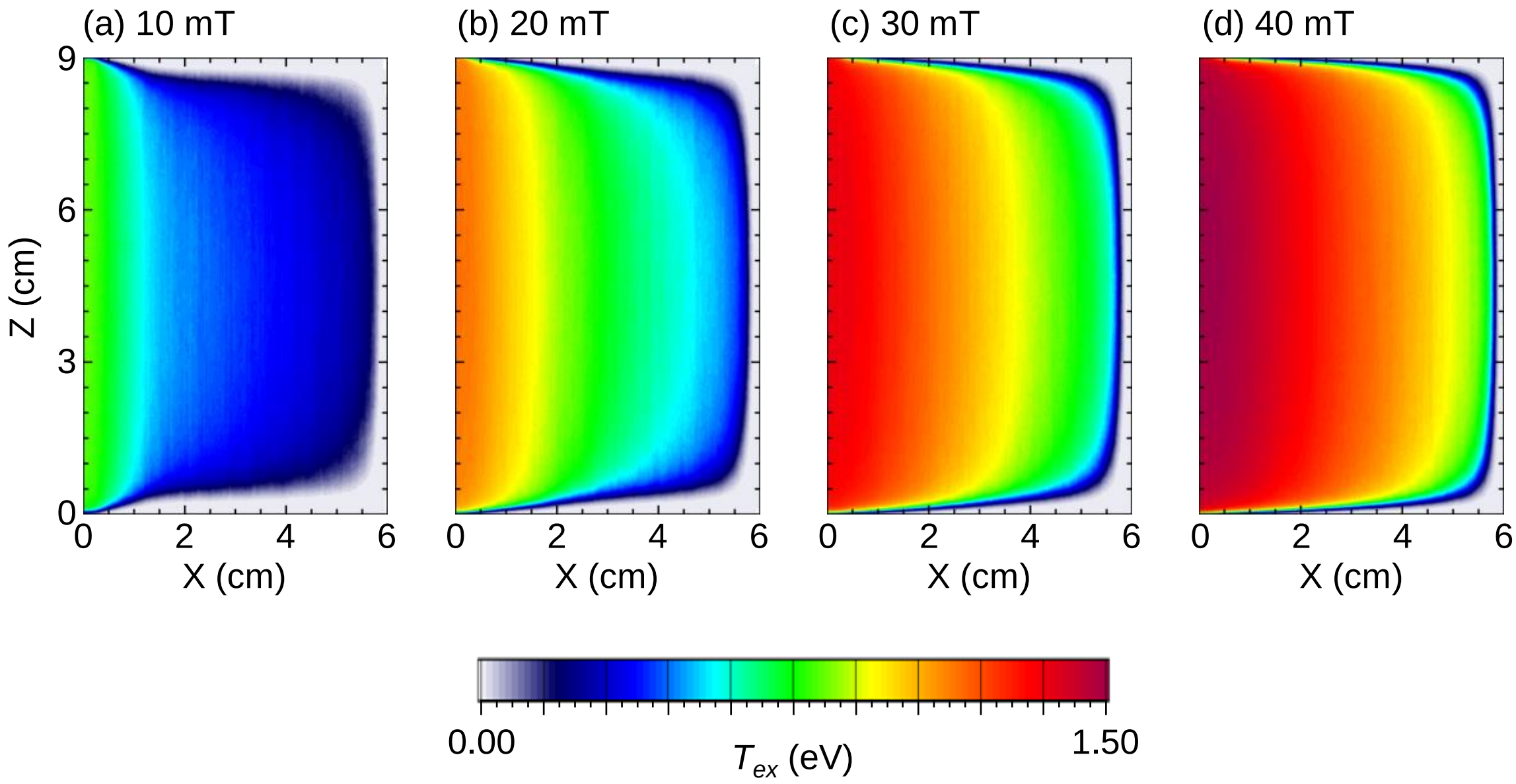

Figure 9
Rauf *et al.*

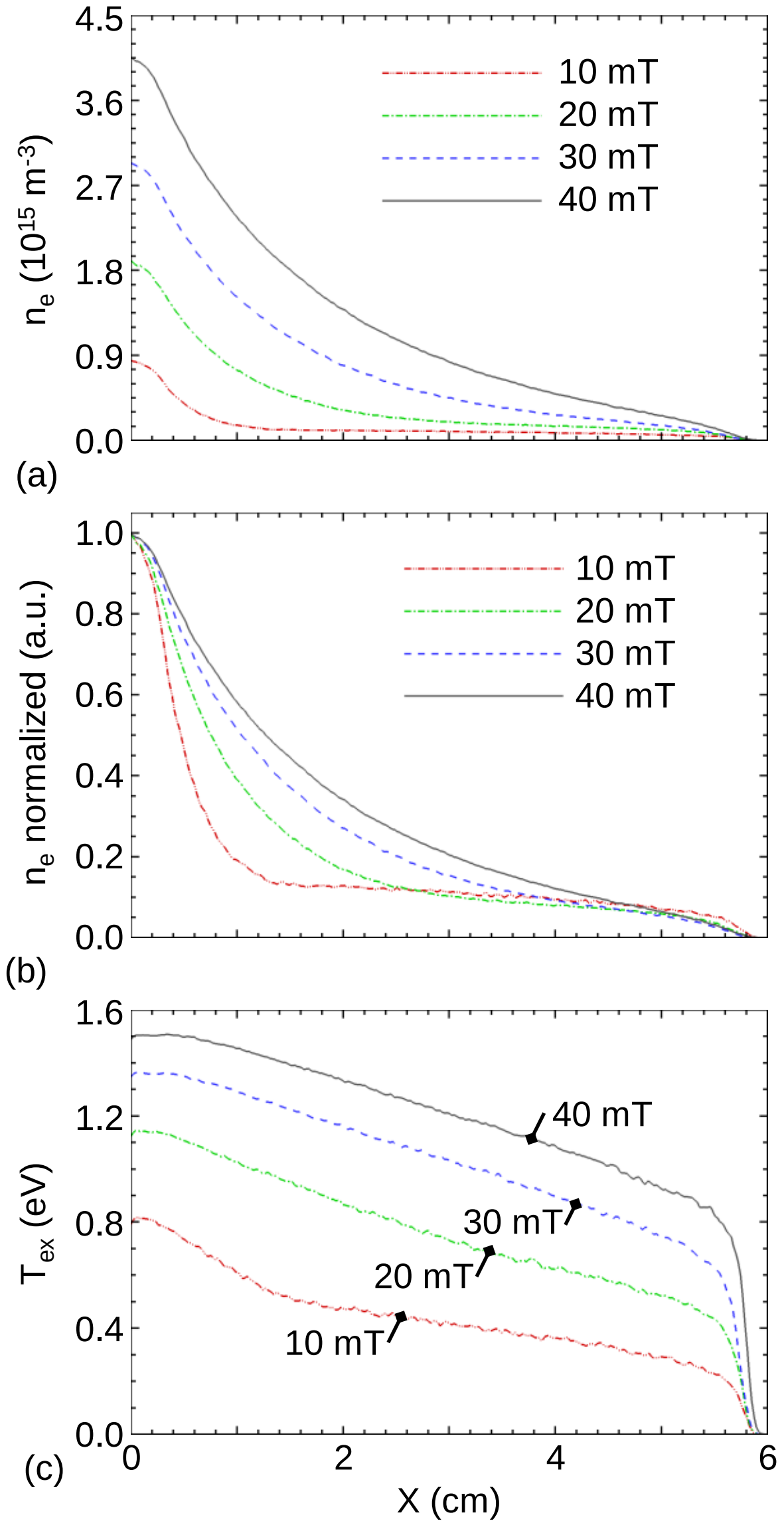

Figure 10
Rauf *et al.*

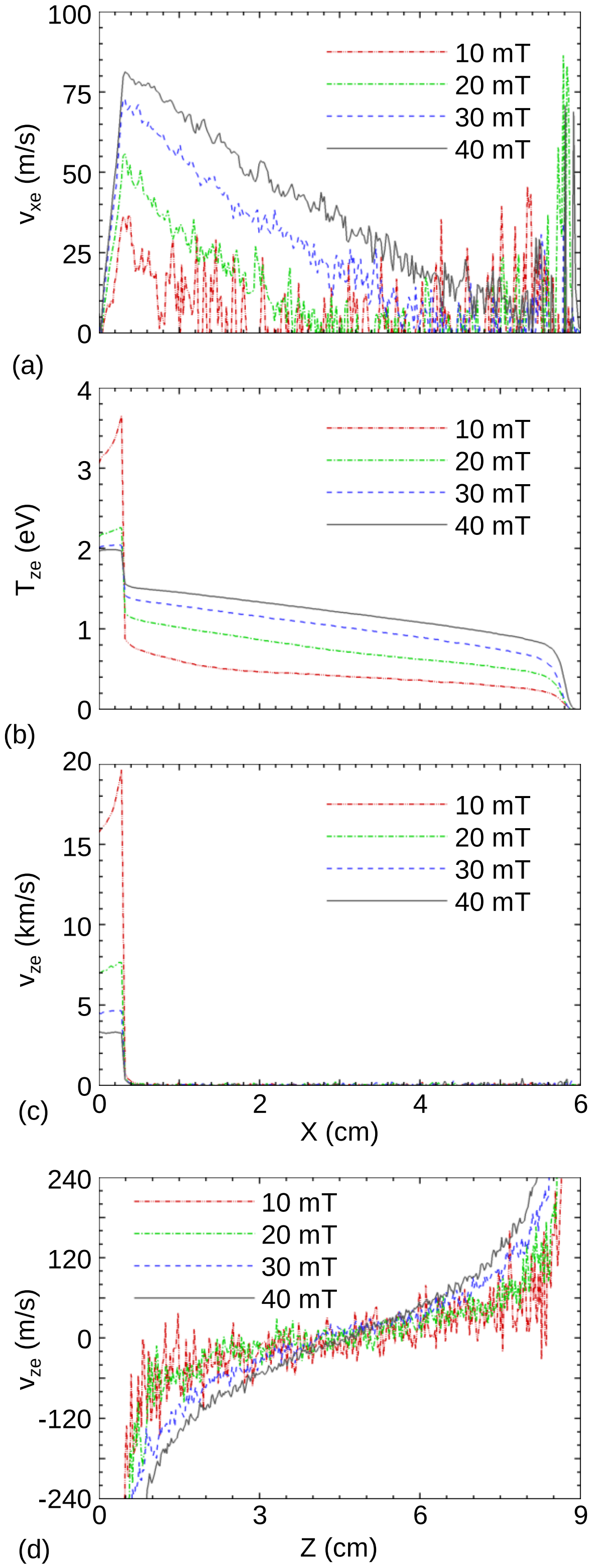

Figure 11
Rauf *et al.*

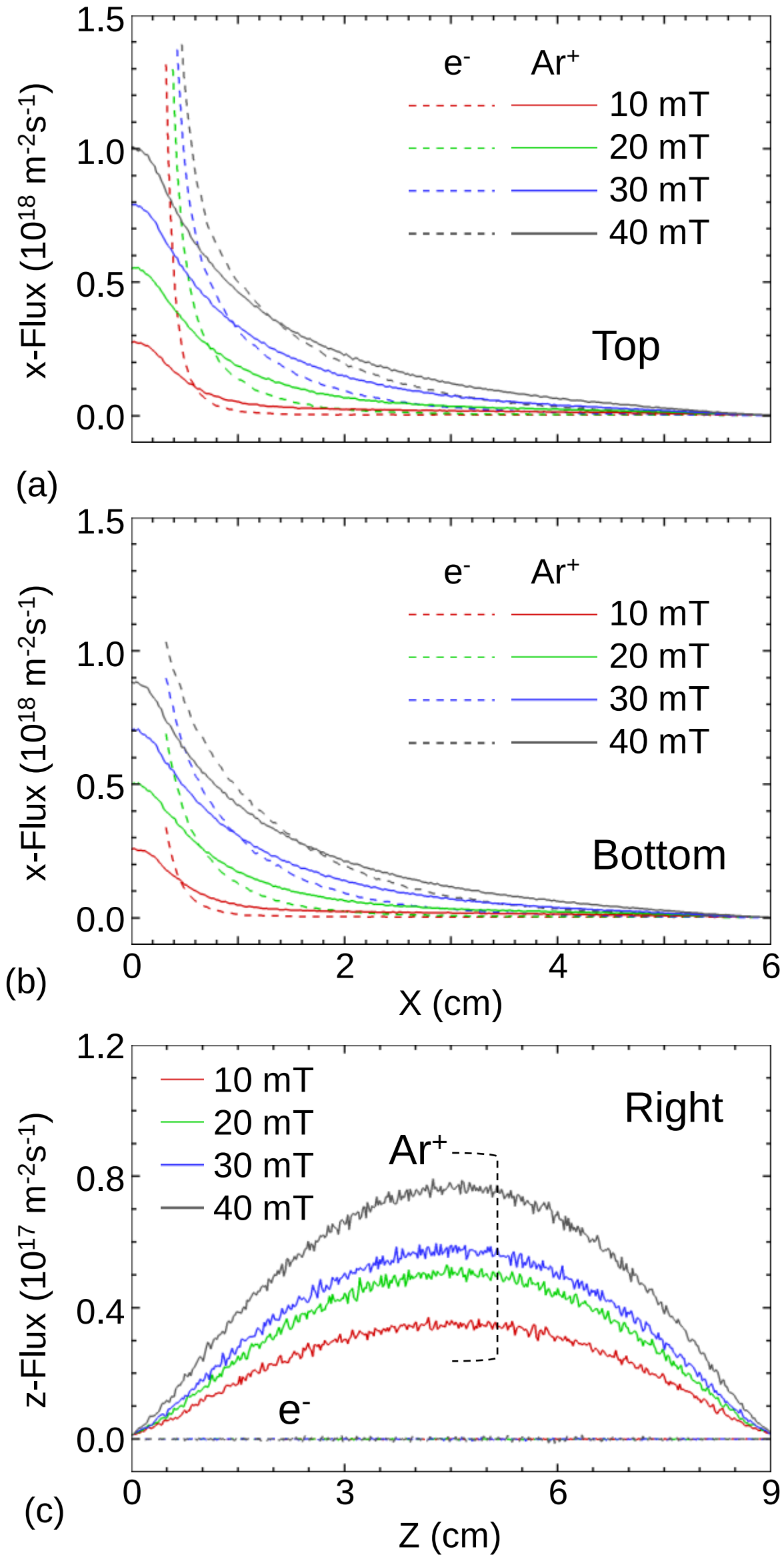

Figure 12
Rauf *et al.*